\documentstyle[epsfig]{elsart} 
\begin{document}
\begin{frontmatter}

\setlength{\baselineskip}{0.85cm}

\title{A Three-Dimensional Calculation of Atmospheric Neutrino Fluxes$^1$}


\author{Yaroslav Tserkovnyak$^2$}
\address{Physics Department, Harvard University, Cambridge MA 
02138}

\author{Robert Komar, Christian Nally, Chris Waltham$^3$}
\address{Department of Physics and Astronomy, University of British
Columbia,
Vancouver B.C., Canada V6T 1Z1}

\thanks[NSERC]{The work was supported by the Natural Science and
Engineering Research Council of Canada (NSERC).}

\thanks[Yaro]{Supported in part by a Faculty of Science scholarship 
from
the University of British
Columbia and a
summer studentship from the TRIUMF laboratory, Vancouver}

\thanks[corr]{Corresponding author; email: waltham@physics.ubc.ca}

\begin{abstract}

We present a three-dimensional calculation of atmospheric neutrino
fluxes using accurate models of the geomagnetic field, hadronic
interactions, tracking and decays. Results are presented for the
Super-Kamiokande (SK) and Sudbury Neutrino Observatory (SNO)
sites using the GEANT-FLUKA hadronic code. We make a comparison with 
previous one-dimensional
calculations and other three-dimensional results. The recently reported 
geometrical 
enhancement of low energy, horizontal neutrinos is confirmed. The effect
on observed leptons is however, restricted to a small overall increase in
flux. We report east-west asymmetries for each of the neutrino species.

\end{abstract}

\end{frontmatter}

Revised draft for submission to Astroparticle Physics


\newpage

\section{Introduction}

Atmospheric neutrinos are produced constantly by the interaction of
primary cosmic rays with nuclei of the earth's upper atmosphere. These
primary interactions produce pions and kaons which decay into neutrinos
and muons; the muons themselves may decay into additional detectable
neutrinos
if they
don't have enough energy to reach the ground. The following reaction
sequence is typical:

\begin{equation}
p+ ^{14}\mbox{N} \rightarrow \pi^+(K^+) + X; \ \ 
\pi^+(K^+) \rightarrow \mu^+ \nu_\mu; \ \
\mu^+ \rightarrow e^+ \bar\nu_\mu \nu_e
\end{equation}

The primary cosmic rays are
mostly protons, with a small fraction of heavier nuclei. The energy
spectrum has the  approximate form $E^{-2.7}$ over the energy range
applicable to the production of detectable neutrinos. The spectrum is
modified at low energies by solar activity, which varies slowly with time.
At large distances from the earth the primary flux is isotropic.
Closer to the earth the flux is reduced at low energies in an
angle-dependent way by the geomagnetic field.

Atmospheric neutrino fluxes are such that the rate of detection is of the
order 100 per kiloton of detector per year. They were the first naturally
occurring neutrinos to be observed, some thirty years
ago\cite{Gaisser_book}. Since then they
have generated interest, first as a critical background in nucleon-decay
searches, and more recently in their own right as the
means
by which neutrino oscillations were discovered\cite{ANP}.  

The most obvious predictions of the reaction sequence above is that (a)
there will be two muon neutrinos for each electron neutrino, (b) the
antineutrino to neutrino ratio will be about unity for muons and 1.3 
($\approx \pi^+/\pi^-$ ratio) for electrons.
In addition to the numerology, we can also make a broad statement about
neutrino isotropy. At high energies the primary cosmic ray flux becomes
isotropic in angle,
as geomagnetic effects become small. It is therefore expected that
the angular distribution of atmospheric neutrinos will tend at high energy
to isotropy, except for a horizontal enhancement due to the increased
meson decay path in the atmosphere.

Prediction (a) has not been
borne out by recent experiments which measure a muon to
electron ratio which is both a factor two too small and which is strongly
zenith angle (and therefore pathlength) dependent\cite{ANP}. This is
evidence for neutrino
oscillations and a
non-zero neutrino mass. Prediction (b) has not yet been tested; there has
been 
no measurement of the antineutrino
to neutrino ratios.
Given the tremendous implications of this neutrino mass signal,
it is plainly important to model atmospheric neutrinos as well as
possible, and examine very closely any mundane effect which may affect the
result.  There have already been several successful calculations of
atmospheric neutrino fluxes. Most have been one-dimensional ``slab" models
with geomagnetic cut-offs, chosen for calculational simplicity. This
approximation is based on well-founded principles and not thought to
introduce serious error, except in the angular distributions at the lowest
detectable energies. We chose to pursue a three-dimensional calculation,
this
route now being feasible because available computing power has increased
dramatically since the first atmospheric neutrino codes were started.

There are three major independent 1-D calculations of atmospheric neutrino
fluxes which predate this work. Two of them were being updated until very
recently; that of the Bartol group
(``BGS"\cite{BGS})  and that of the (Super-)Kamiokande group
(``HKHM"\cite{Honda}). Bugaev
and Naumov (``BN") is an older work \cite{BN}. Lee and Koh \cite{LK}
(``LK") was an early attempt to make the BGS code three-dimensional; it
had some problems and was not pursued \cite{Gaisser}.
The main justification for the 1-D approximation is that the transverse
momentum from pion
or muon decay is reckoned to be small enough (10s of MeV/c) to obviate the
need to consider more dimensions when the threshold of the atmospheric
neutrino detectors is at least 100 MeV/c.

As our work was being done, Battistoni {\it et al.}\cite{Battistoni}
published the first results from a 3-D calculation. They reported a
geometrical enhancement of low energy, horizontal neutrinos, not seen
in 1-D calculations. The 
origins
of this effect were subsequently explained in a pedagogical paper by
Lipari \cite{Lipari_geom}. At the time we did not see this effect because
we had expanded our detector size for calculational efficiency and had
unwittingly washed out the enhancement. Reducing the vertical extent of
the detector revealed the effect to be there.  The impact on the observed
leptons is however, restricted to a small overall increase in flux and a
small but significant improvement in understanding the so-called
``east-west effect" seen in the Super-Kamiokande (SK)
data\cite{Futagami,Lipari_ew}. Other than this, neither our work nor 
Battistoni's
report a large difference between the 1-D and 3-D approaches.

In this work we compare our calculations with those of
Bartol 1-D model\cite{BGS} for the SK site, those of the
3-D model of Battistoni\cite{Battistoni} for SK and northern sites,
and with newer 3-D work of Wentz\cite{Wentz} and the HKKM\cite{HKKM} 
group.

In section 2 the details of the calculation are described. 
We outline the input parameters for the primary flux, the geomagnetic 
field, and the atmosphere. We discuss tracking issues and dealing with 
secondary particles. The question of hadronic interactions is considered 
in some detail, with comparisons made between various codes.
Lastly we outline the findings of other 3-D work on angular 
distributions. 

In section 3 we present our results 
for the SK and Sudbury 
Neutrino Observatory (SNO)\cite{SNO} sites and compare them with
other calculations. The 
magnetic latitude for SK 
is 
27$^\circ$~N\cite{HKKM}, and that for SNO is 
57$^\circ$~N.

\section{Theory}

\subsection{Introduction}

In the absence of magnetic fields and at high energy, the production of 
neutrinos by cosmic
ray interactions in the earth's atmosphere can be calculated to a fair
approximation by semi-analytic models\cite{Gaisser,Volkova}. This is
especially true at high energies, where the primary proton (or heavier
nucleus), intermediate meson, and decay product muons and neutrinos,  are
all essentially co-linear. The mesons - mostly pions and kaons - have a 
choice of interaction or decay in the atmosphere, and the muons - if of
low enough energy - can also decay. The cosmic ray primaries are
isotropic; the only deviation from isotropy in the neutrinos is a
horizontal enhancement due to a large slant range in the atmosphere,
allowing more to decay.

\begin{eqnarray}
P^{\pi,K,\mu}_{\nu_\mu+\bar\nu_\mu}
\sim E^{-\gamma} \nonumber
\cdot
\left({{1}\over{1+\frac{6 E\cos\theta}{\mbox{\small
121GeV}}}}+
{{0.213}\over{1+{{1.44 E \cos\theta}\over{\mbox{
\small 
897GeV}}}}}\right)(\mbox{m}^{-2}\cdot\mbox{sr}^{-1}\cdot\mbox{s}^{-1}
\cdot\mbox{GeV}^{-1})
\label{Volkova spectrum}
\end{eqnarray}

This expression, due to Volkova\cite{Volkova}, gives the muon neutrino and
antineutrino flux $P^{\pi,K,\mu}_{\nu_\mu+\bar\nu_\mu}$ which results from 
$\pi$, $K$ and $\mu$ decay. It is
good
to a few \% at energies above 100~GeV and angles not close to horizontal
($|\cos\theta| > 0.2$) where 
the curvature of the earth becomes important. 
We do not use this formula in our calculation, but it encapsulates the
basic features of the neutrino flux.
At lower energies the neutrino spectrum follows that of the
primaries, $E^{-\gamma}$, where $\gamma\approx2.7$, and is isotropic.
At higher energies the neutrino spectrum goes as
$E^{-(\gamma+1)}/\cos\theta$ due to competition between interaction and
decay in the atmosphere.
The first term in the parenthesis accounts for pion decay, the second,
kaon decay. At low energies one has to account for the geomagnetic field,
muon decay, and non-colinear decay products.


\subsection{The Basics of Atmospheric Neutrino Flux Calculations}

There are two basic approaches to these calculations; they are either
one-dimensional (``slab") models or three-dimensional models like the one
presented here.  The calculations can be split into different parts, some
of which are common between the two approaches.

\begin{itemize}

\item 3-D: Start at the top of the atmosphere close to the location of a
detector and take primaries with energies according to the interstellar
spectrum but only above the local cutoff (which depends on the direction
of the primary). 1-D calculations only need primaries in direct line of
sight to the detector.
Check if earth shadowing prevents this trajectory by reversing
charge of primary and tracking backwards. If this track intersects the
earth's surface this primary is rejected. 

\item Allow the primaries to interact with a model of the atmosphere,
using one or more packages of hadronic interaction codes,
producing
charged secondaries and neutrinos.
\item  Track the charged secondaries in the earth's magnetic field (3-D).
In the 1-D model the secondaries are tracked along the primary direction.
\item Allow secondaries to decay into neutrinos. 
\item Check if the neutrino has crossed the designated detector area
(inflated in such a way as to improve statistics without washing out local
effects). 
\item Bin neutrinos and calculate fluxes as a function of neutrino type,
energy, direction and solar activity.
\item Calculate interaction rates and angular distributions of leptons
observed in 
the detectors.
\end{itemize}

The statistical accuracy and number of bins is typically limited by
computation time in the 3-D case. 

\subsection{Primary Flux}
\nopagebreak

For the primary cosmic ray flux we use the same
parameterization as the Bartol
calculation
\cite{Agrawal},
and assume
medium solar activity. 
This form has been 
superseded by a new fit by the Bartol group (Gaisser {\it et 
al.}\cite{Gaisser2001}) which predicts a 10\% increase in the 
primaries which produce sub-GeV neutrinos.

\subsection{Geomagnetic Field}
\nopagebreak

The geomagnetic field is calculated using a 10th
order
multipole expansion, with  spherical harmonic
coefficients
taken from the IGRF 1995 model \cite{IGRF}. 
Errors in the field at the earth's surface are less than 25 nT ($<$0.1\%). 
A detailed discussion of geomagnetic effects appears in a paper by
Gaisser {\it et al.} \cite{Gaisser}.


For any given position on the earth's surface, an angular map may be made
showing the minimum rigidity required for a positively charged particle,
starting from a large distance away, to reach the surface from that angle.
The resulting cutoff maps for SNO and SK are shown in 
Fig.\ref{fig:cutoff}; in terms of 
cardinal points the x-axis is ordered
N-W-S-E-N.
The plot for the SK site is identical
to that obtained by Honda et al. \cite{Honda}. There is much less
structure
in the more northerly SNO site, with the cutoffs from the north being much
lower than those from the south (the opposite of the SK case), and very
little of the east-west asymmetry seen at SK\cite{Futagami}.
The maps can be  
understood in 
terms of the
angle of the geomagnetic field as seen from the detector position: the
positions of the SNO and SK sites are shown in Fig.\ref{fig:geomag1}.
Because the dip angle at the SK site is small, there is a strong deficit
of primaries coming from the east. At the SNO site, the dip angle is
large, so the east-west asymmetry is small, and there is almost no
reduction in primaries coming from above. However, where the line-of-sight
intersects the geomagnetic field at 90$^\circ$, there is an increase in
the cutoff. This occurs for near-horizontal primaries, and for those
coming from a tilted ring around middle latitudes in the southern
hemisphere.

Figure \ref{fig:pr} shows how the geomagnetic field affects the spectrum 
of
cosmic ray primaries reaching the upper atmosphere.
Maps of the intensities of cosmic ray
primaries on the earth's surface (averaged over all incident angles and
shown as a fraction of the zero geomagnetic field case) are shown for
various momenta.
  The effect of the geomagnetic
field is of course strongest at low energies and fades away by about 
20~GeV at the SNO site, and 40~GeV at the SK site. 
Note how the SNO site sits on the edge of a
region where the low energy primary flux is changing very rapidly 
with latitude.
This provides a challenge for
the 3-D modeller as the detector cannot be arbitrarily enlarged to improve
statistics.

\subsection{Tracking Primaries and Secondaries}

Far away from the earth, where the geomagnetic field merges into the
interplanetary field, the flux of cosmic rays is
essentially isotropic\cite{isotropy}. 
As a cosmic ray approaches the earth, it is progressively bent by the
geomagnetic field and, if it is of low enough rigidity (momentum/charge)
it will be turned away. A typical value is that required for a
particle to orbit the earth,  just above the surface, at its magnetic
equator, with a locally horizontal field of mean strength 30~$\mu$T; this
rigidity is 60~GV/$c$ (i.e. an energy of 60~GeV for a proton).
In addition, some parts of the earth's surface
are, for certain angles and rigidities, in shadow from other parts of the
earth.

We model these effects in a manner similar to that of the HKHM group
\cite{Honda}. 
The primary flux at large distances from the earth only depends on energy
(and time - due to a small solar modulation, which we ignore for now) and
can be written as $\phi^\infty_p(E)$. The primary flux which strikes the
earth's atmosphere, however, depends on position and angle: 
$\phi_p(E,{\bf x},\Omega)$.
Invoking Liouville's theorem, which is applicable as
long
as the magnetic field can be considered static, yields the simple form:

\begin{eqnarray}
\phi_p(E,{\bf x},\Omega) & = & \phi^\infty_p(E) \mbox{ for allowed paths}\\
			    & = & 0  \mbox{ for forbidden paths}
\end{eqnarray}  

To decide which paths are allowed or forbidden, 
a check is made for reflection in the geomagnetic field
and shadowing
by the earth before a positively charged primary is tracked
through the
atmosphere.
An equivalent negatively  charged
particle is tracked
backwards to see if its trajectory reaches a very large distance
from the earth (10 $R_E$, where $R_E$ is the earth's
radius) without intersecting with 
it. If it did,
such a trajectory is allowed; if not, it is rejected.

The
atmospheric density is calculated according to Linsley's compilation of 
 data\cite{LSA}. 
Inside the atmosphere both primaries and
secondaries were tracked in steps of 1~km (Fig. \ref{fig:3d_an}). In each 
case helical steps were
taken, using the magnetic field at the centre of the helix and assuming it
to be uniform over the step. In order to keep statistics sensible,
detector areas were artificially increased to a rectangle 10$^\circ$ by
40$^\circ$ with the shorter side pointing to magnetic north to reduce
the washing out of local geomagnetic effects. The ``detector" was
considered to be a flat sheet as any increase in the thickness was found
to mask the geometrical horizontal enhancement discovered by Battistoni
{\it et al.}\cite{Battistoni}.

\subsection{Hadronic Interactions}
\label{section:hi}

In order to calculate hadronic interactions, we start with experimental
parameterization for hadron-hadron cross-sections (from the
CORSIKA\cite{Corsika} air shower package). Hadron-nucleus and
nucleus-nucleus cross-sections were calculated using Glauber theory
(instead of $A^{2/3}$ re-scaling as used in GEANT).  

Secondaries were generated using several standard hadronic packages.
We used generators CALOR, GEANT-FLUKA and GHEISHA within GEANT
3.21\cite{Geant,GF},
and stand-alone generator 
FRITIOF and decay routine JETSET\cite{Fritiof}. 
In
order to compare the results from these packages, we calculate
several functions of the pion production cross sections in the energy
range relevant to sub-GeV neutrinos. In this we followed the published
comparisons of Gaisser et al. \cite{Gaisser}. The crucial discriminants
are the  spectrum-weighted moments, the ``Z"-factors:

\begin{equation}
Z_{p\rightarrow\pi^\pm}=\int^1_0 dx\cdot x^{1.7} 
\frac{dn_{\pi^\pm}(x,E_N)}{dx} 
\end{equation} 

where $x=E_\pi/E_N$, and $E_N$ is the total energy of the incident nucleon
in the lab system and $E_\pi$ is the energy of the secondary pion.
The values of these factors determine how many charged pions are produced
at a given energy, and it is these pions which decay into neutrinos.
In Fig.\ref{fig:Zfactor} we plot these moments for charged pions for three
hadronic
packages. The GEANT-FLUKA\cite{GF}
package gives the highest value, and
GHEISHA
\cite{Geant} 
the lowest. In the middle lie CALOR\cite{Geant} and FRITIOF\cite{Fritiof},
although FRITIOF gives the highest multiplicities.
It is a combination
of these (CALOR below 10~GeV/c and FRITIOF above), which most closely
reproduces the spectrum-weighted moments in the TARGET code of
BGS\cite{Gaisser}. CALOR and GEANT-FLUKA are not independent; the two
codes are identical above 10~GeV.
CALOR uses
GEANT-FLUKA pion and nucleon routines phased in from 2 to 10~GeV.
Kaon and antinucleon routines in the two codes are identical.
In terms of neutrino production the three hadronic packages 
produced results 
with very similar distributions but with different 
overall
rates.
GEANT-FLUKA and CALOR/FRITIOF give
similar overall rates while GHEISHA leads to a much smaller neutrino flux 
(about
40\% of the rate calculated with GEANT-FLUKA).


The quantity which indicates best the neutrino yield is the
spectrum-weighted production of charged pions
$Y(\pi^\pm)E_0^{-1.7}$, where $Y$ is the multiplicity per interaction and
$E_0$ the primary energy.
In Fig.\ref{fig:gaisser_comp}
we compare the spectrum-weighted pion production per nucleon
for 
CALOR/FRITIOF and for TARGET. The pion
momentum range
of 3-4~GeV/c is chosen as these are the pions which produce neutrinos
with energies around 1~GeV\cite{Gaisser}.  
The value of this parameter integrated on a logarithmic scale
in the primary energy is roughly
proportional to the measurable neutrino flux. 
Due in part to different pion multiplicities, our CALOR/FRITIOF
combination gives an
integral which is 21\% lower than TARGET. Hence we expect that differences
in hadronic codes, everything else being equal, 
will give us a  
neutrino flux at 1~GeV which is about 21\% lower than that of BGS when we
use CALOR/FRITIOF hadronic code. Furthermore, GEANT-FLUKA
produces somewhat smaller unweighted charged pion multiplicities than
FRITIOF, with consequently smaller neutrino fluxes. See Section
\ref{section:fluxes} below.

\subsection{East-West Effect}

A major systematic effect on atmospheric neutrino data is the geomagnetic
field. This is particularly true at more southerly sites like SK, where
the most pronounced feature is an east-west asymmetry. 
To understand this asymmetry lends significant 
confidence
in the interpretation of the data. As Lipari has pointed
out\cite{Lipari_ew}, there is a hierarchy in the asymmetries of the
different neutrino species, due to the difference in the bending of
primaries and
secondaries (Fig. \ref{fig:east_west}). The $\nu_e$ arises from the decay 
of positively charged
secondaries and so the asymmetry is the largest. The $\bar\nu_e$ arises
from
the decay of negatively charged
secondaries and so the asymmetry is the smallest. In between lie the
$\bar\nu_\mu$ and $\nu_\mu$, with the former having a larger asymmetry as
some of them arise from decaying positive muons which are more deflected
than
neutrinos produced directly from pion decays. 

The effect is not easy to find in the data without energy and directional
cuts.
Lipari proposes a using neutrinos which produce leptons with energies of 
0.4-3~GeV with zenith angles $|\cos\theta_l <0.5|$. 
The standard
definition of the asymmetry $A$ in terms of the numbers of neutrinos
$N_{E,W}$ is as follows.

\begin{equation}
A=\frac{N_E-N_W}{N_E+N_W}
\end{equation}

With a 
3-D calculation
he obtains an electron E-W asymmetry of 0.224 and a muon asymmetry  of
0.091. 
The corresponding experimental values\cite{Futagami} are in good 
agreement, being $0.21\pm0.04$ and $0.08\pm0.04$ respectively.

\subsection{Horizontal Enhancement}

The first three-dimensional calculations revealed a geometrical effect 
not anticipated in one-dimensional models. For low energy primaries 
($\ll$1~GeV) the 
secondary particles have almost no directional correlation with their 
parents (Fig. \ref{fig:lipari_effect}). For a detector far from the 
earth's centre,  this produces a 
strong enhancement for particles detected coming from near the horizontal
\cite{Battistoni,Lipari_geom}. Unfortunately this effect is very hard to 
observe for three reasons: (a) the flux of low energy secondary neutrinos 
is small, (b) the cross sections for interaction in a detector is small,
and (c) the correlation between the directions of the neutrino and the 
detected lepton is weak.

\section{Results}

Our results using the GEANT-FLUKA hadronic interaction code are shown in 
the last six plots of this paper. The spectrum,
azimuthal and zenith angle distributions are shown in Figs. 
\ref{fig:sk-3},\ref{fig:sk-1},\ref{fig:sk-2} for the SK site, and  Figs. 
\ref{fig:sno-3},\ref{fig:sno-1},\ref{fig:sno-2} for the SNO site.
In these figures, the points and error bars are our data, the higher sets 
of lines
are from the BGS 1-D calculation\cite{BGS}, and the lower sets are from 
the 3-D
calculation of Battistoni\cite{Battistoni}.
The results for each site arise from  back-tracking $22\times10^6$ protons 
and 
$2.5\times10^6$ 
alphas. The number of neutrinos crossing each detector is  $3.1\times 
10^6$.

\subsection{Fluxes}
\label{section:fluxes}

In overall fluxes, our results are in good agreement with Battistoni 
{\it et al.}, who use the same primary flux and a similar hadronic
generator. The BGS\cite{BGS} results are somewhat higher, largely due to 
differences in hadronic generators as noted above.


\subsection{East-West Effect}

The azimuthal angle distributions for the SK site are shown in
Fig.\ref{fig:sk-1} and those for SNO are in Fig.\ref{fig:sno-1}.
After making the cuts proposed by Lipari\cite{Lipari_ew}, numerical values
are given in table \ref{table:ew}.
Our model reproduces the correct hierarchy of East-West asymmetries, in
order of highest to lowest: $\nu_e, \bar\nu_\mu, \nu_\mu, \bar\nu_e$.
The lepton asymmetries are calculated by weighting the neutrino values
with the quasi-elastic scattering cross sections\cite{LS} and convoluting
with the neutrino lepton scattering angle. The mean value of this angle 
at these energies is 36$^\circ$\cite{Futagami}. The much smaller effect 
of detector 
resolution is 
not taken into account; the same is true for Lipari's work.
However, our lepton asymmetries are significantly higher than those of 
Lipari and
SK data; the overall electron asymmetry is $\sim$0.32, and the muon 
asymmetry is 
 $\sim$0.23.
We note that a recent 3-D calculation of Wentz {\it et al.} \cite{Wentz}
also reports large asymmetries. This needs further investigation.

\subsection{Horizontal Enhancement}

The zenith angle distributions for the SK site are shown in
Fig.\ref{fig:sk-3} and those for SNO are in fig.\ref{fig:sno-3}.
The energy range is 0.5~GeV$<E_\nu<$3~GeV.
Our calculations are presented with those of Bartol (1D) and 
Battistoni
(3D). Lipari's horizontal enhancement can be seen in the 3D models,
especially at the SNO site where neutrino spectrum is much softer.
Honda {\it et al.} find a 10\% smaller enhancement which they ascribe
to their use of a different hadronic interaction model.
This feature is not present in the 1D calculation. However, it is almost
completely washed out in the detection of neutrino-induced leptons, as
demonstrated in Refs. \cite{Battistoni,HKKM}.

\section{Conclusions}

This work describes a full three-dimensional
simulation of atmospheric neutrino fluxes. It largely
justifies previous and widely used (1-D) approximations. 

We note
that the overall fluxes depend on our choice of hadronic code. We
can discount GHEISHA which is known not to work too well at the low
energies 
important in generating quasi-elastic leptons. That leaves CALOR/FRITIOF 
and GEANT-FLUKA, which produce very similar spectra, angular 
distributions,
and
fluxes.

We confirm the geometrical horizontal enhancement.
The east-west asymmetries predicted by our model are higher than those
of Lipari\cite{Lipari_ew} and SK data\cite{Futagami}, but similar to those 
of 
Wentz\cite{Wentz}.

\section*{Acknowledgements} 

The authors would like to thank E. Kearns and C. Walter for the
use of the Boston University Physics Department's O2000 CPU farm. We 
also thank T. Gaisser
and P. Lipari for useful discussions. The work was supported by the
Natural Science and
Engineering Research Council of Canada (NSERC). One of us (YT) is grateful
for a Faculty of Science scholarship from the University of British
Columbia, and a
summer studentship from the TRIUMF laboratory.

\newpage

\begin{center}

\begin{table}[h]
\begin{tabular}{|c|c|c|c|c|c|}
\hline\hline
Species & SK (data) & SK (Lipari) & SK (This Work) & SNO (Lipari) 
& SNO (This Work) \\
\hline\hline
$\nu_e$ &  & 0.335 & $0.390\pm0.006$ & 0.214 & $0.161\pm0.015$\\ \hline
$\bar\nu_e$ & & -0.065 & $0.158\pm0.010$ & -0.061 & $-0.034\pm0.027$\\
\hline
$\nu_\mu$ & & 0.028 & $0.223\pm0.006$ & 0.002 & $0.033\pm0.010$ \\ \hline
$\bar\nu_\mu$ &  & 0.240 & $0.320\pm0.005$ & 0.153 & $0.094\pm0.009$ \\
\hline
\hline
$e$ & $0.21\pm0.04$ & 0.224 & 0.32 & 0.136 & 0.11  \\ 
\hline
$\mu$ & $0.08\pm0.04$ & 0.091 & 0.23 & 0.046 & 0.05 
\\ \hline
\hline \hline
\end{tabular} 
\caption{
East-west asymmetries for each neutrino and lepton species, for the SK and
SNO sites. Neutrinos are selected with $E$ = [0.5,3.0]~GeV (leptons
with [0.4,3.0]~GeV)
and
$\cos\theta < 0.5$. The SK data are from Ref. \cite{Futagami}. 
Lipari's values are from Ref. \cite{Lipari_ew}.
 }
\label{table:ew}
\end{table}

\end{center}

\newpage

\section*{Figure Captions}
\noindent

\begin{figure}
\center{\epsfig{file=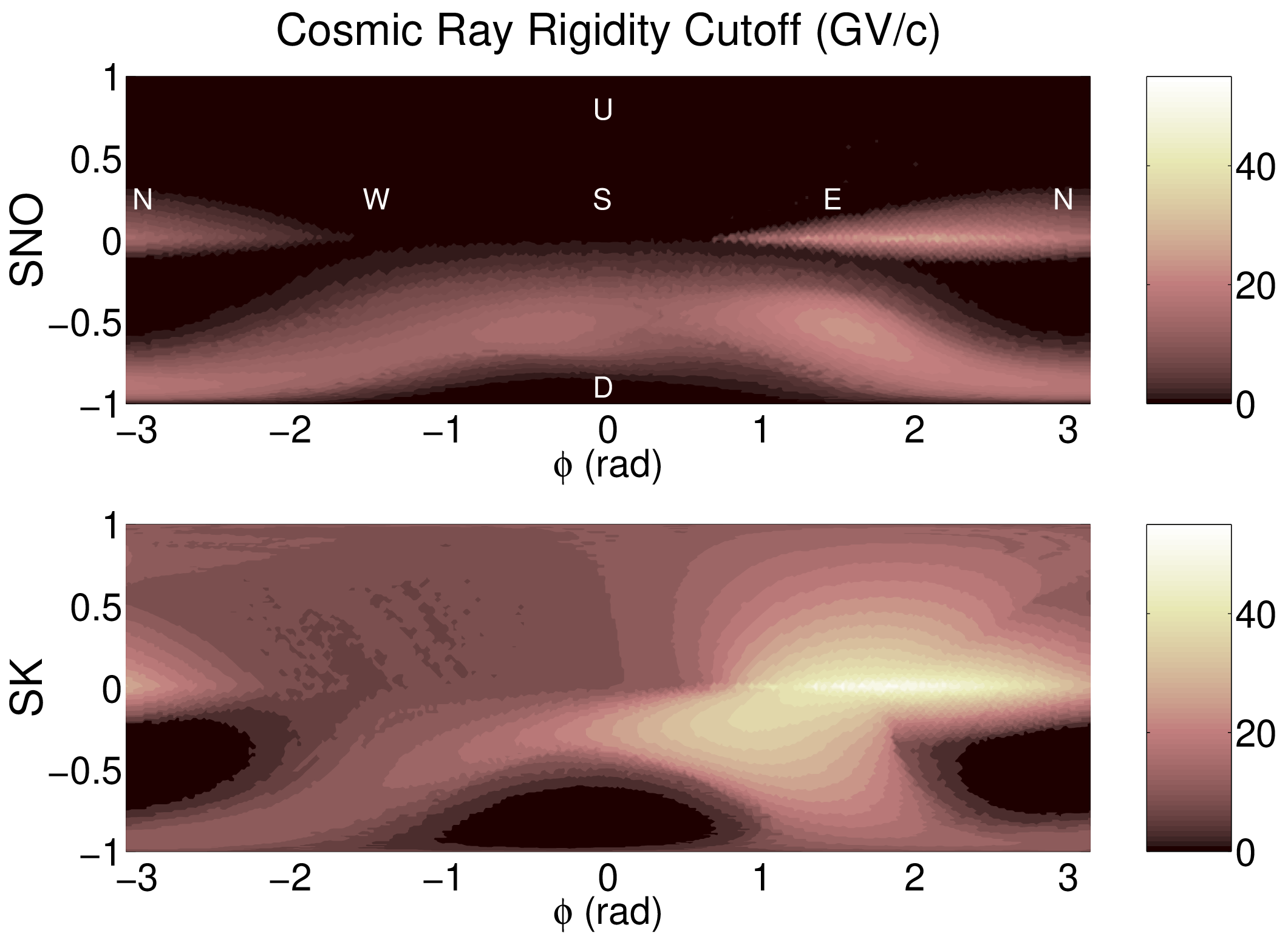,width=12cm,angle=90}}
\vspace*{1cm}
\caption{}
\label{fig:cutoff}
\end{figure}

Figure \ref{fig:cutoff}: Rigidity cutoffs in the 1-D model for SNO (top)
and SK
(bottom) sites. The more northerly SNO
site has in general much lower cutoffs. The SNO site has the opposite
up-down asymmetry to 
the SK site,  and a very much smaller east-west asymmetry. These cutoffs
were not used in the calculation, they are merely means to check part of
the  code, and as an aid to understanding. In terms of cardinal points the
x-axis is ordered N-W-S-E-N.

Figure \ref{fig:geomag1}: The cutoff maps of Fig.\ref{fig:cutoff} can be
understood in terms of the angle of the geomagnetic field as seen from the
detector position.  Because the dip angle at the SK site is small, there is
a strong deficit of primaries coming from the east (see
Fig.\ref{fig:geomag1}. At the SNO site, the dip angle is large, so the
east-west asymmetry is small, and there is almost no reduction in
primaries coming from above. However, where the line-of-sight intersects
the geomagnetic field at 90$^\circ$, there is an increase in the cutoff.
This
occurs for near-horizontal primaries, and for those coming from a tilted
ring around middle latitudes in the southern hemisphere.

\begin{figure}
\center{\epsfig{file=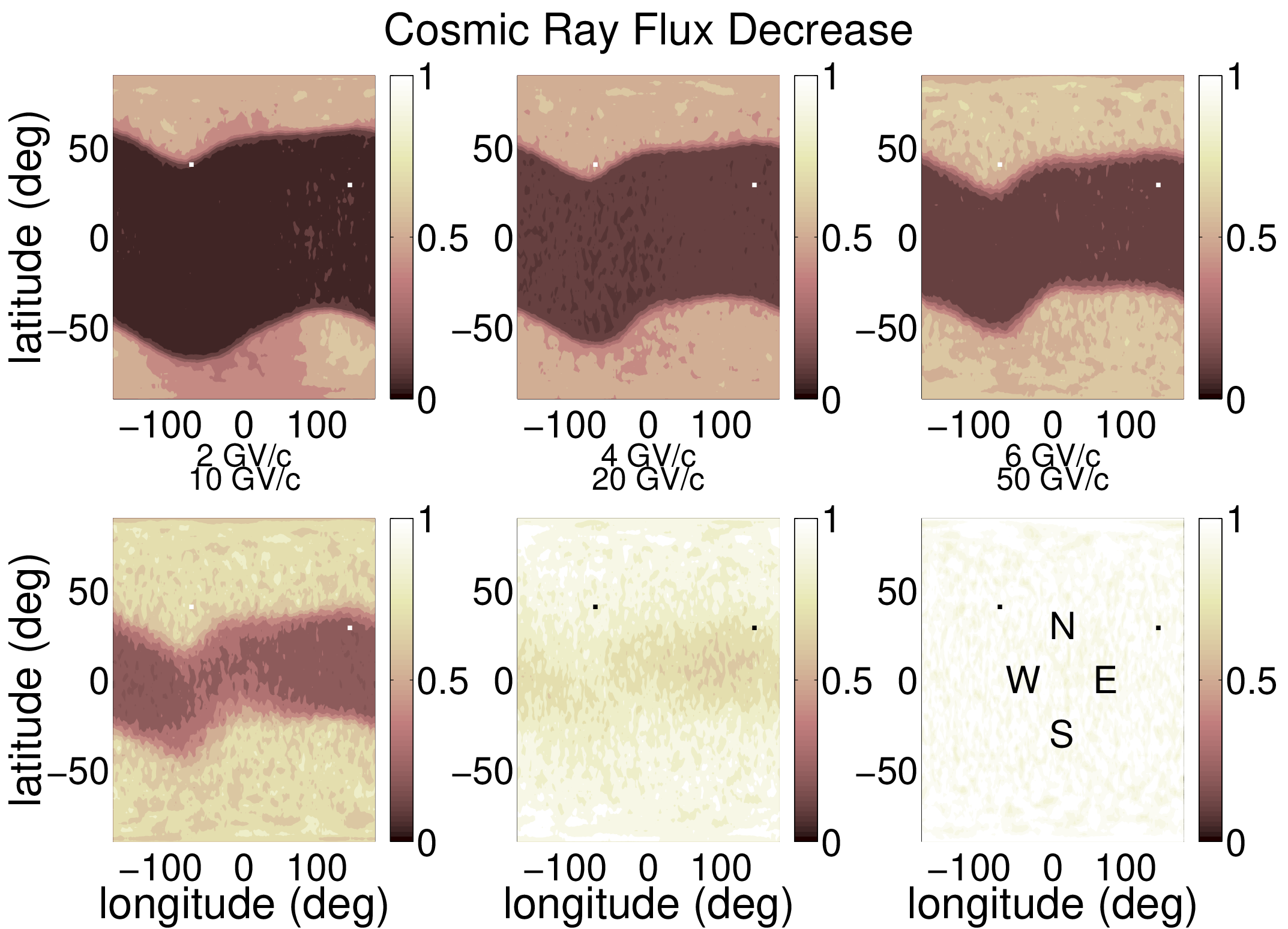,width=13cm,angle=90}}
\vspace*{1cm}
\caption{}
\label{fig:pr}
\end{figure}

Figure \ref{fig:pr}: 
Maps of the intensities of cosmic ray primaries on
the earth's surface (averaged over all incident angles and shown as a
fraction of the zero geomagnetic field case) are shown for 2,4,6,10,20 and
50~GeV/c. The two small white crosses mark the positions of SNO and
SK.

\begin{figure}
\center{\epsfig{file=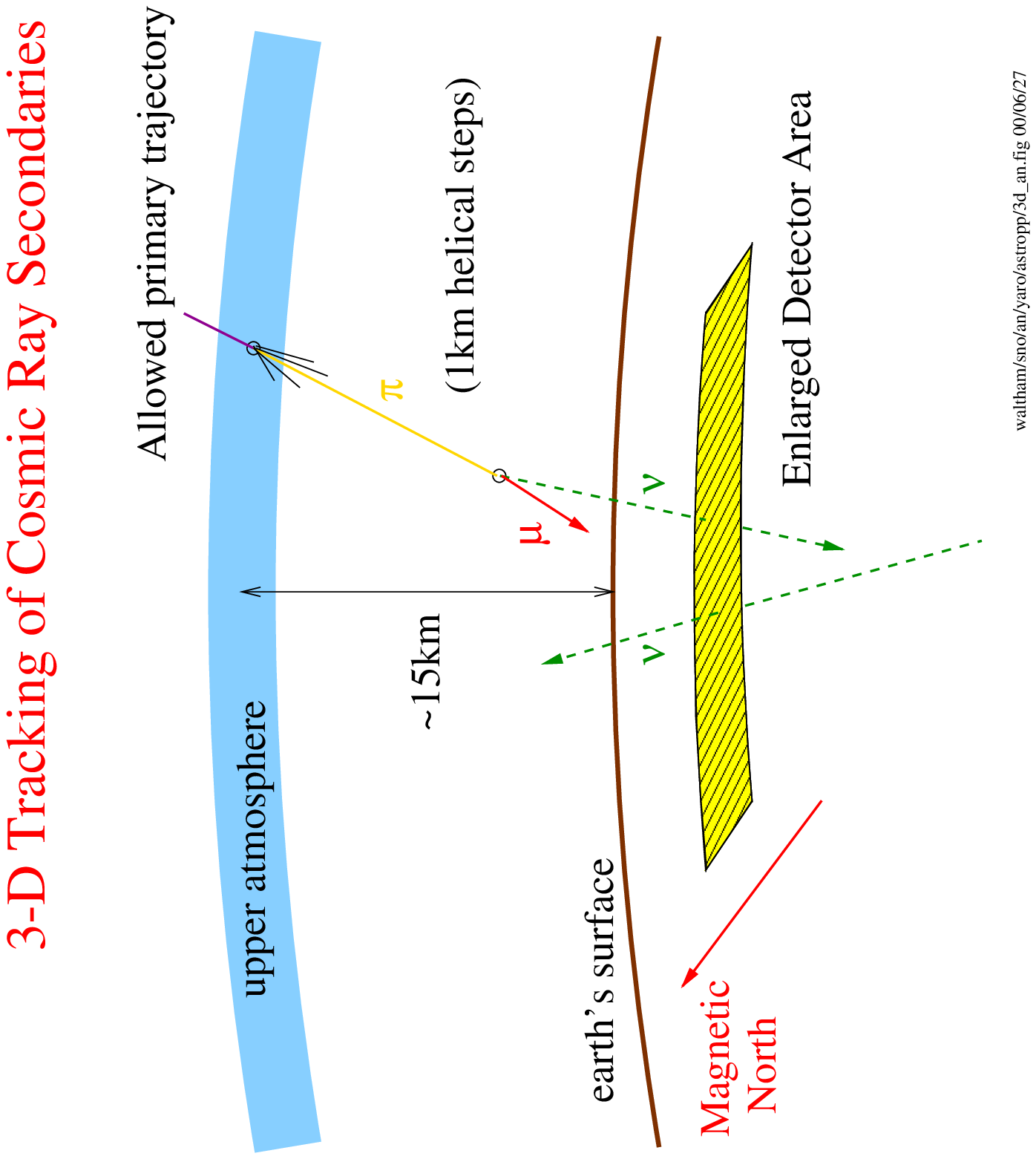,width=12cm,angle=-90}}
\vspace*{1cm}
\caption{}
\label{fig:3d_an}
\end{figure}

Figure \ref{fig:3d_an}:  Sketch of the proton + air nucleus $\rightarrow
\pi \rightarrow \mu$ production and decay chain which leads to muons being
observed deep underground in SNO. Pions decay in flight in the atmosphere
to produce observable muons and neutrinos.

\begin{figure}
\center{\epsfig{file=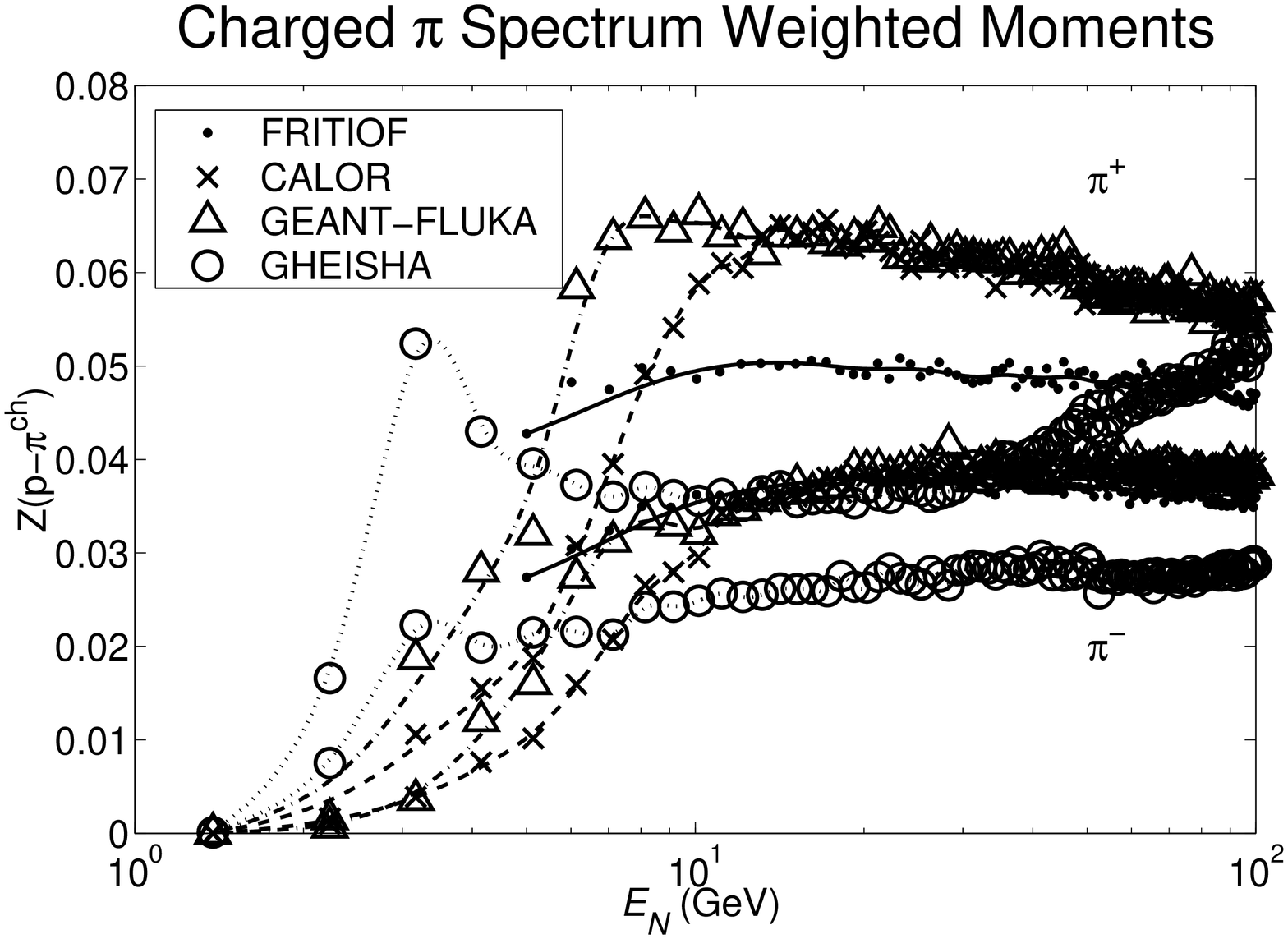,width=11cm,angle=90}}
\vspace*{1cm}
\caption{}
\label{fig:Zfactor}
\end{figure}

Figure \ref{fig:Zfactor}: Spectrum-weighted moments for the production of
charged pions as a function of proton total lab energy, for different
common hadronic codes.
The moment value are calculated by
GHEISHA (low; o$\cdot \cdot \cdot$o$\cdot \cdot \cdot$o),
CALOR (X - - - X - - - X), FRITIOF ($\cdot$---$\cdot$---$\cdot$)
 and
GEANT-FLUKA (high; $\triangle$-$\cdot$-$\triangle$-$\cdot$-$\triangle$).
For each calculation the higher of the two data sets 
represents $\pi^+$ production, and the lower, $\pi^-$. 
The main results of this work are obtained using GEANT-FLUKA.
The combination with a
spectrum-weighted moment
which closely resembles that in
BGS\cite{Gaisser} is  CALOR
below 10~GeV/c
and
FRITIOF above 10~GeV/c.

\begin{figure}
\center{\epsfig{file=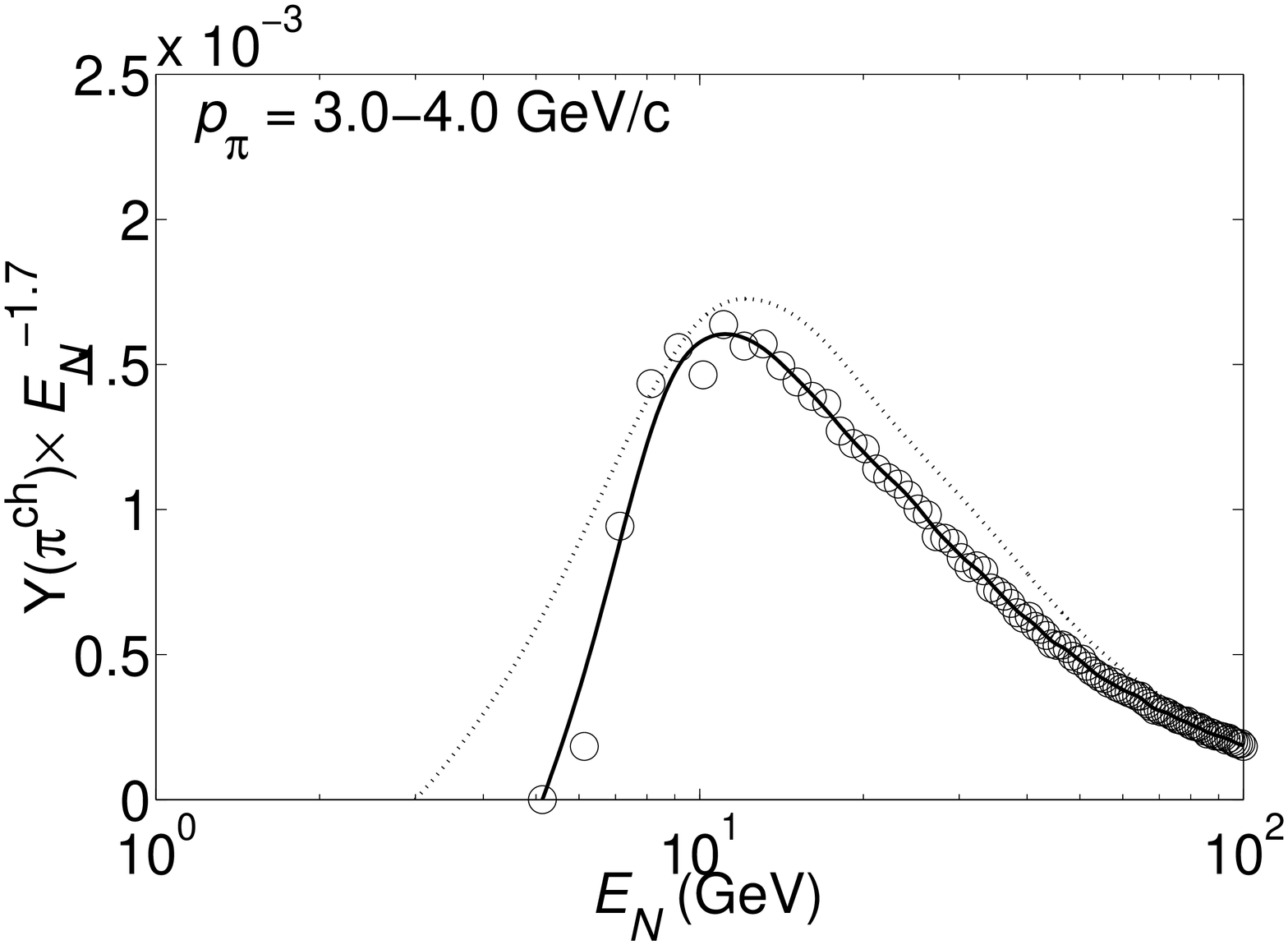,width=11cm,angle=90}}
\vspace*{1cm}
\caption{}
\label{fig:gaisser_comp}
\end{figure}

Figure \ref{fig:gaisser_comp}:
Spectrum-weighted production of charged
pions of 3-4~GeV/c as a function of primary energy. This momentum bin is
responsible for neutrino production with energies around 1~GeV. This work
(CALOR/FRITIOF: o---o---o) is compared with BGS ($\cdot \cdot
\cdot \cdot \cdot$)
\cite{Gaisser}.

\begin{figure}[h]
\center{\epsfig{file=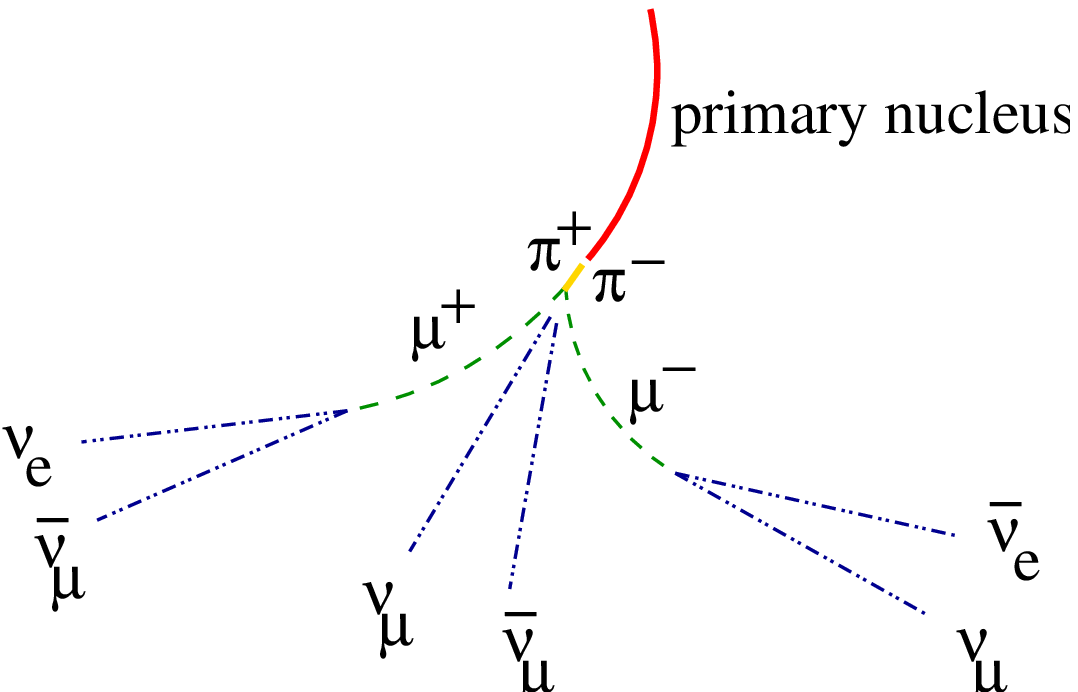,width=13cm,angle=0}}
\vspace*{1cm}
\caption{}
\label{fig:east_west}
\end{figure}

Figure \ref{fig:east_west}: This diagram shows how the relative bending of 
 charged primaries and secondaries contributes to the east-west asymmetry 
of the various neutrino species. The hierarchy of asymmetries is expected 
to be (high to low): $\nu_e, \bar\nu_\mu, \nu_\mu, \bar\nu_e$.

\begin{figure}[t!]
\center{\epsfig{file=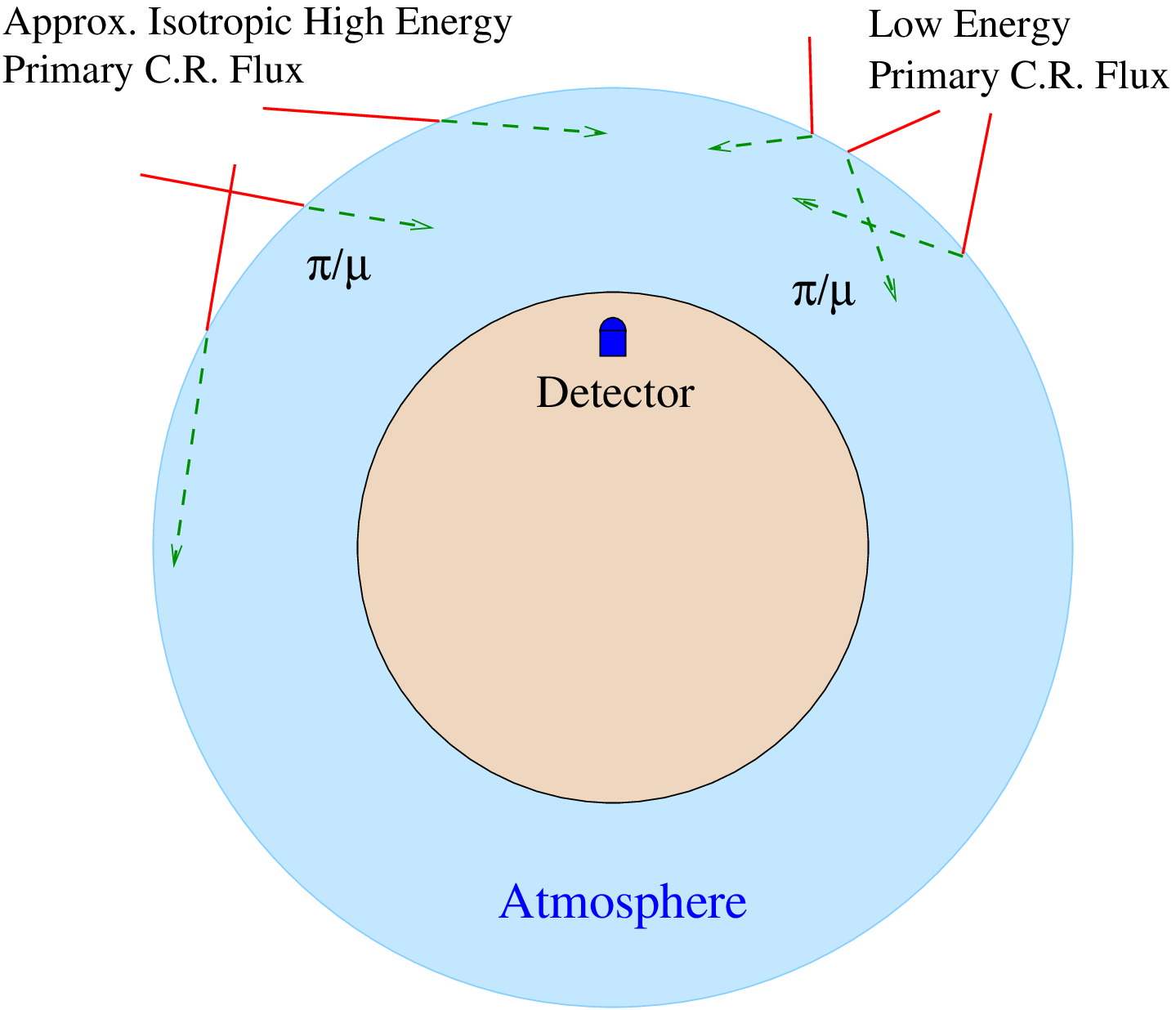,width=13cm,angle=0}}
\vspace*{1cm}
\caption{}
\label{fig:lipari_effect}
\end{figure}

Figure \ref{fig:lipari_effect}: High and low energy atmospheric neutrinos 
are expected to have different zenith angle distributions. Both peak at 
the horizontal, but for different reasons. High energy neutrinos come from 
high energy mesons, which are produced collinearly 
with the almost isotropic primaries. However, they  are more likely to 
decay in 
the 
greater path length of atmosphere at the horizontal. In contrast, low 
energy cascades loose all angular correlation with the primary, resulting 
in a purely geometrical enhancement at the horizontal.

Figure \ref{fig:sk-3}: Neutrino energy spectra for SK site. 
The data
points are this work, using the GEANT-FLUKA hadronic code. The lines are 
calculations of other groups. Those with the horizontal peak are from the 
3D model of
Battistoni et al.  \cite{Battistoni}; the solid lines are neutrinos and
the dotted antineutrinos. The lines with the slight horizontal dip are
from the Bartol 1D model \cite{Agrawal}; the dashed lines are  
neutrinos
and the dash-dotted antineutrinos.

\begin{figure}
\center{\epsfig{file=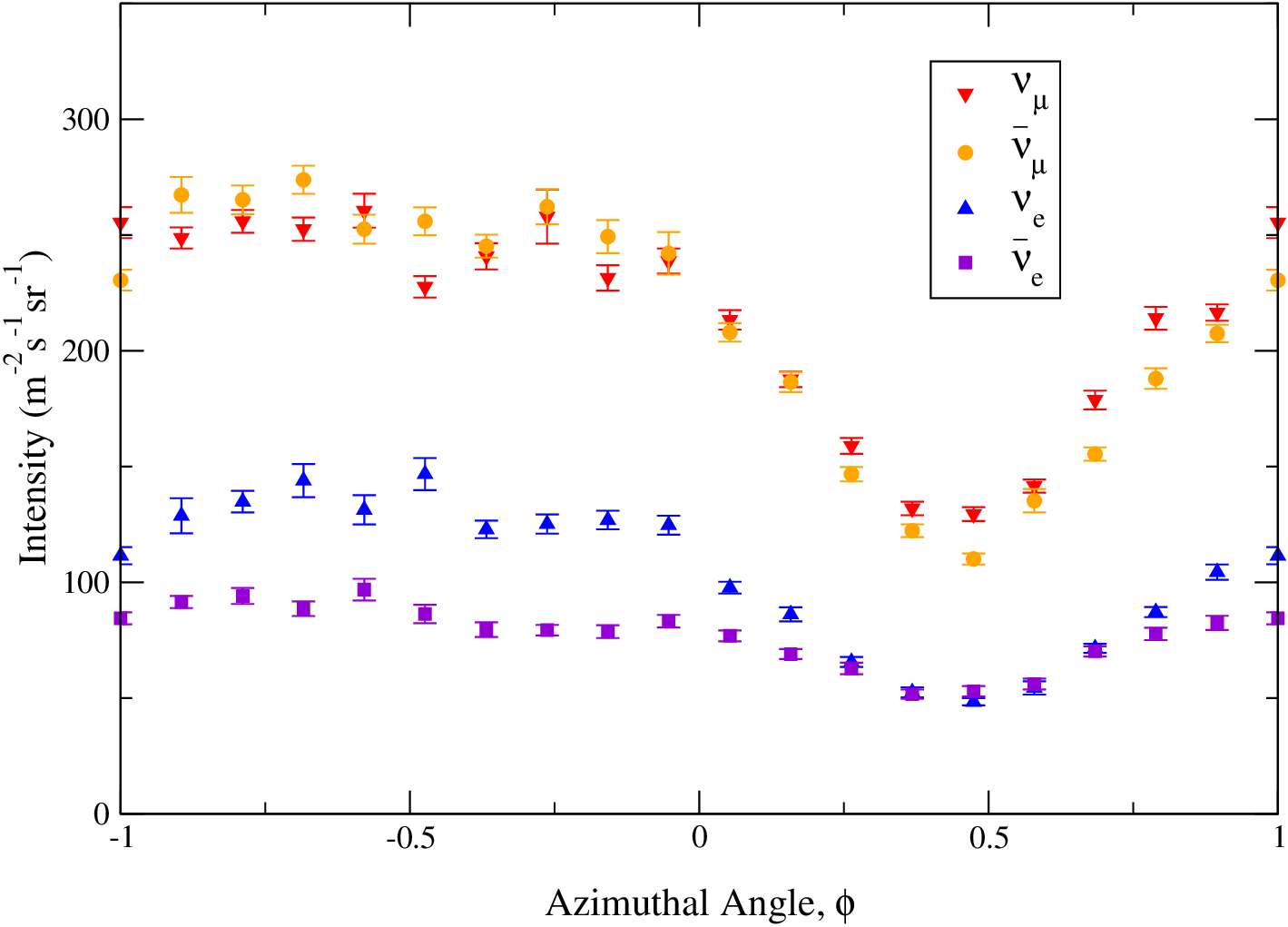,width=15cm}}
\vspace*{1cm}
\caption{}
\label{fig:sk-1}
\end{figure}

Figure \ref{fig:sk-1}: Neutrino azimuthal angle distributions for the SK 
site.

\begin{figure}
\center{\epsfig{file=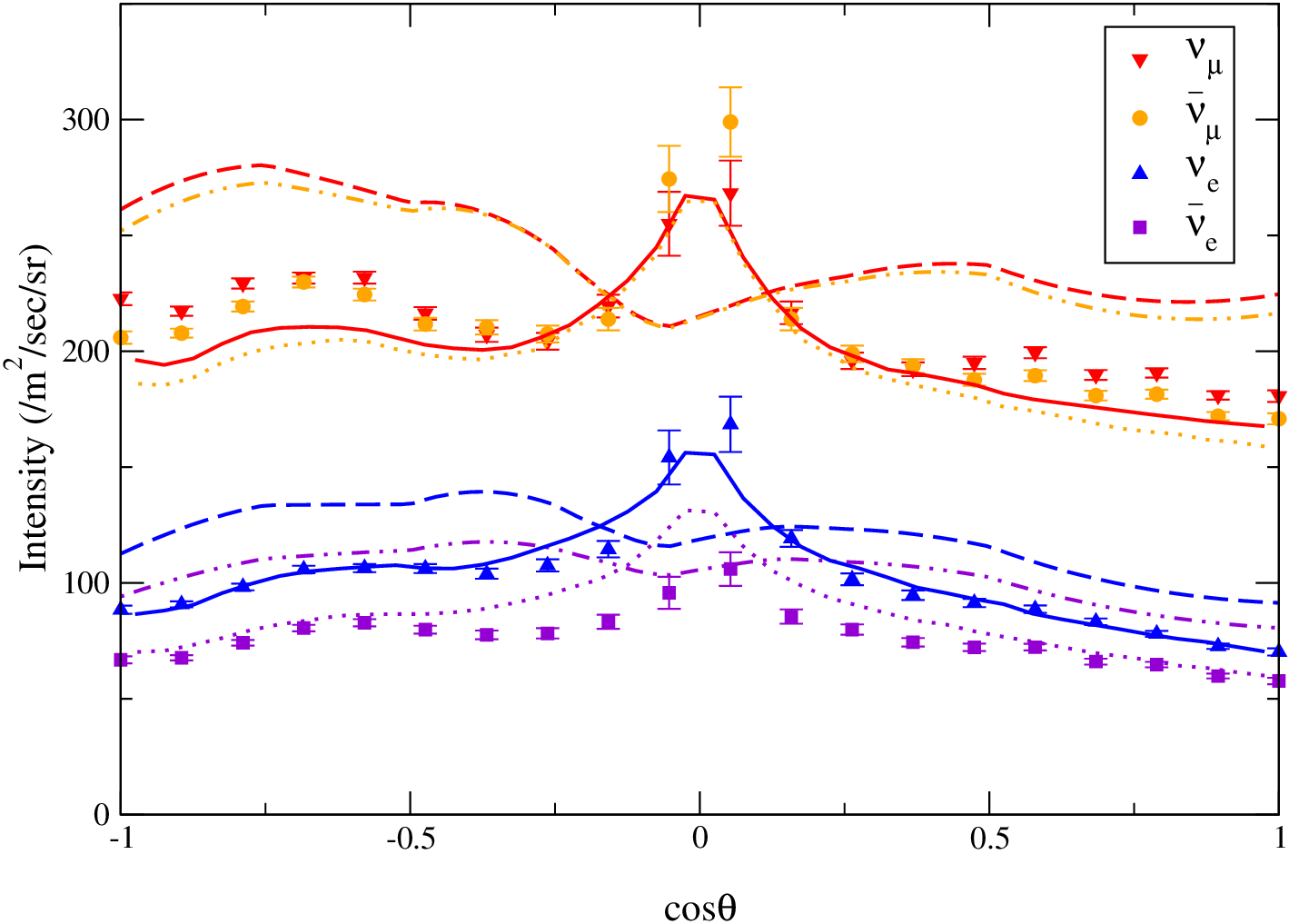,width=15cm}}
\vspace*{1cm}
\caption{}
\label{fig:sk-2}
\end{figure}

Figure \ref{fig:sk-2}: Neutrino zenith angle distributions for the SK 
site.

\begin{figure}
\center{\epsfig{file=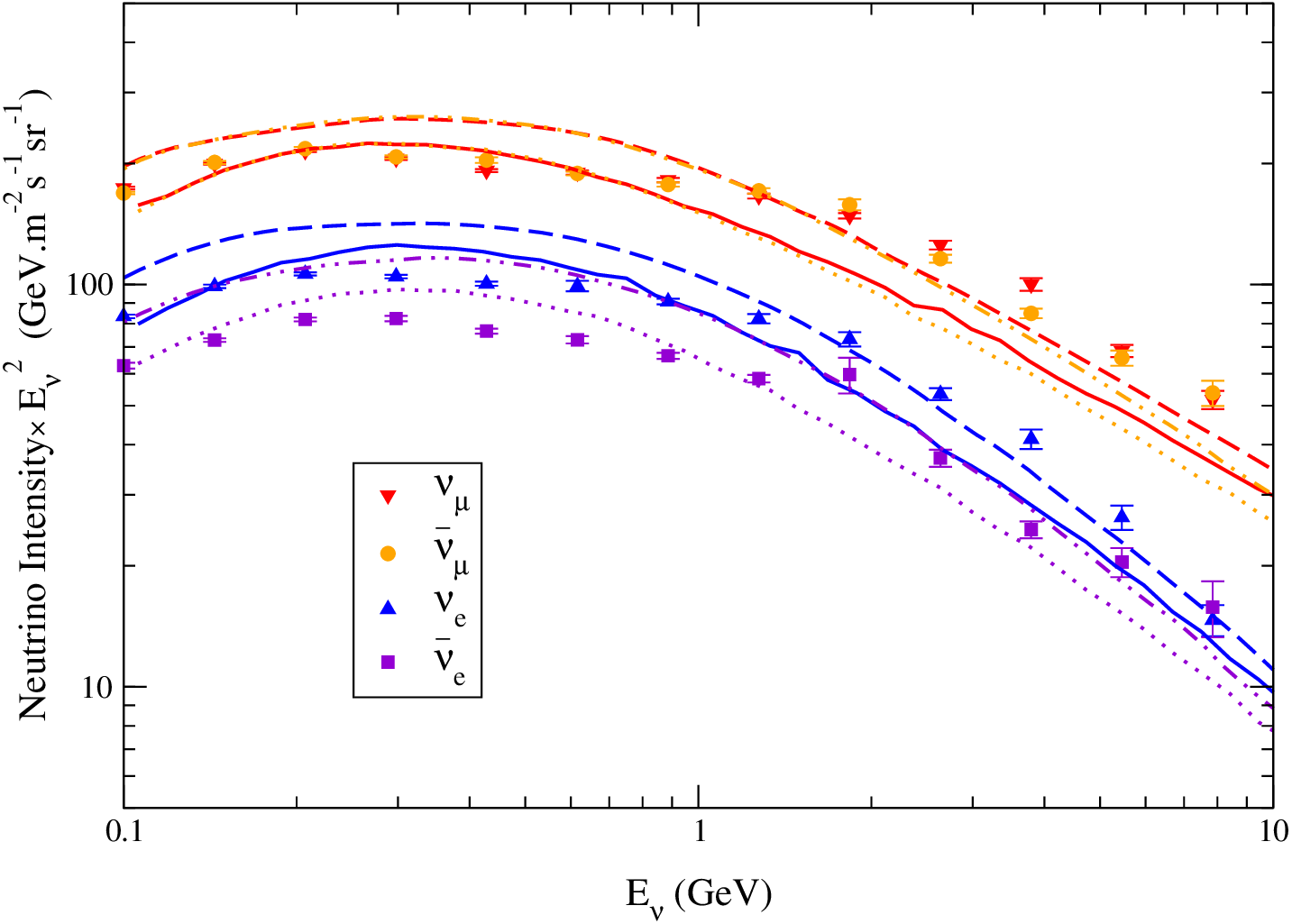,width=15cm}}
\vspace*{1cm}
\caption{}
\label{fig:sno-3}
\end{figure}

Figure \ref{fig:sno-3}: Neutrino energy spectra for the SNO site.
The line
conventions are the same as for SK.

\begin{figure}
\center{\epsfig{file=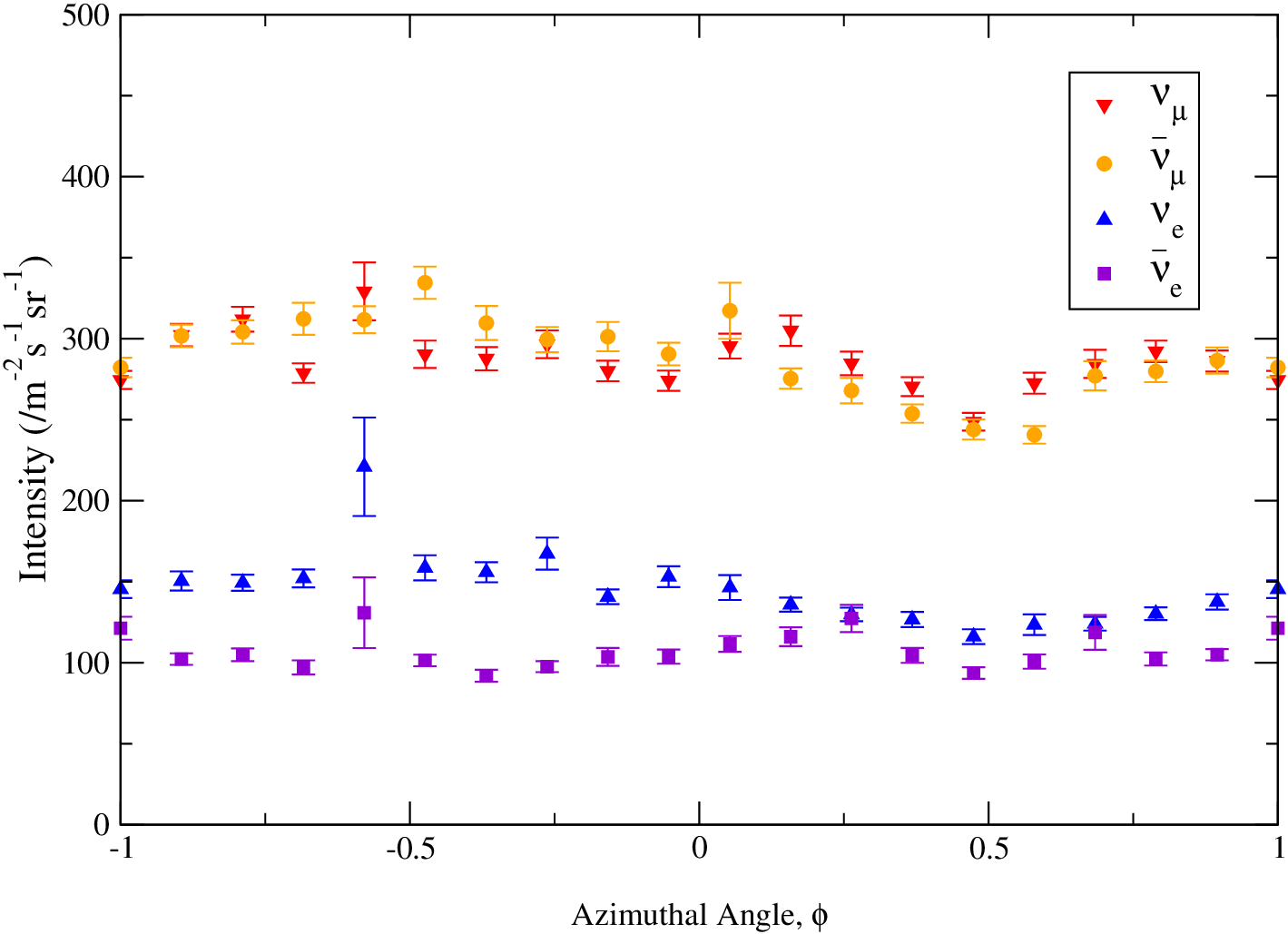,width=15cm}}
\vspace*{1cm}
\caption{}
\label{fig:sno-1}
\end{figure}

Figure \ref{fig:sno-1}: Neutrino azimuthal angle distribution for the SNO 
site. Line
convention as for SK. The
horizontal enhancement is greater at SNO than at SK because the
higher geomagnetic latitude softens the neutrino spectrum.

\begin{figure}
\center{\epsfig{file=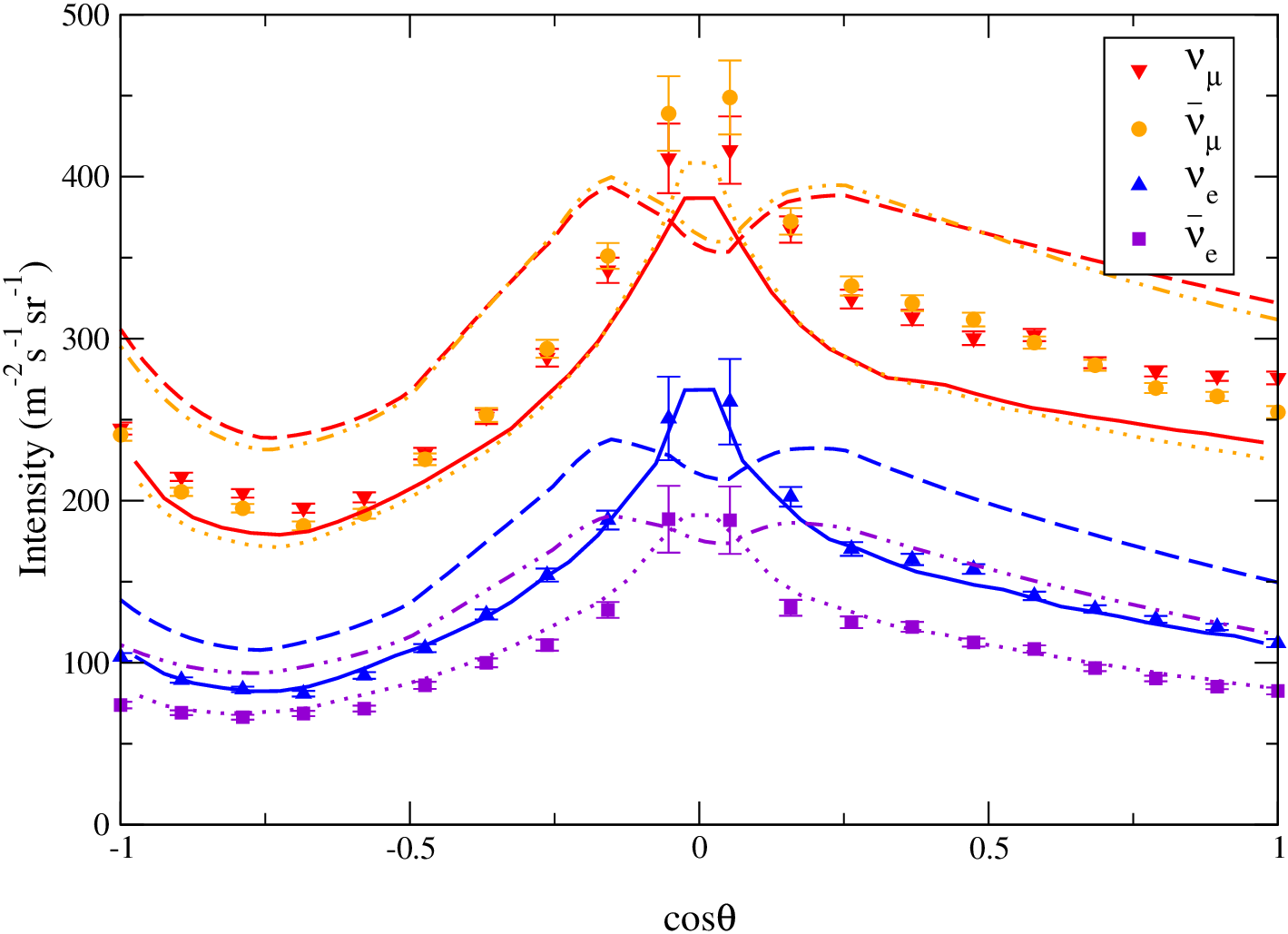,width=15cm}}
\vspace*{1cm}
\caption{}
\label{fig:sno-2}
\end{figure}

Figure \ref{fig:sno-2}: Neutrino zenith angle distribution for the SNO 
site. Line 
convention as for SK. The  
horizontal enhancement is greater at SNO than at SK because the
higher geomagnetic latitude softens the neutrino spectrum.


\newpage

\newpage

\begin{figure}
\center{\epsfig{file=geomag2.eps,width=15cm,angle=0}}
\vspace*{1cm}
\caption{}
\label{fig:geomag1}
\end{figure}

\newpage

\newpage

\newpage

\newpage

\newpage

\newpage

\newpage
\begin{figure}
\center{\epsfig{file=sk-3.eps,width=15cm}}
\vspace*{1cm}
\caption{}
\label{fig:sk-3}
\end{figure}

\newpage

\newpage

\newpage

\newpage

\newpage

\newpage




\end{document}